\documentclass[twocolumn,showpacs,preprintnumbers,amsmath,amssymb]{revtex4-1}
\usepackage{graphicx}
 \usepackage{epstopdf} % needed for figures
\usepackage{dcolumn}   % needed for some tables
\usepackage{bm}        % for math
\usepackage{amssymb}   % for math
\usepackage{amsfonts}
\usepackage{amsmath}
\usepackage{hyperref}
\usepackage{color}
\usepackage{booktabs}

\newcommand{\EQ}{\begin{eqnarray}}
\newcommand{\EN}{\end{eqnarray}}
\newcommand{\EQQ}{\begin{eqnarray*}}
\newcommand{\ENN}{\end{eqnarray*}}

\renewcommand{\t}{^{\mbox{\tiny\sf T}}}

\newcommand{\bremark}{\begin{remark} }
\newcommand{\eremark}{\end{remark}}
\newcommand{\btheorem}{\begin{theorem}}
\newcommand{\etheorem}{\end{theorem}}
\newcommand{\blemma}{\begin{lemma}}
\newcommand{\elemma}{\end{lemma}}
\newcommand{\bassumption}{\begin{assumption} }
\newcommand{\eassumption}{\end{assumption}}
\newcommand{\bcorollary}{\begin{corollary} }
\newcommand{\ecorollary}{\end{corollary}}
\newcommand{\bdefinition}{\begin{definition} }
\newcommand{\edefinition}{\end{definition}}
\newcommand{\bproposition}{\begin{proposition}}
\newcommand{\eproposition}{\end{proposition}}
\newtheorem{remark}{\rm\bfseries Remark}[section]
\newtheorem{definition}{\rm\bfseries Definition}[section]
\newtheorem{theorem}{\rm\bfseries Theorem}[section]
\newtheorem{lemma}{\rm\bfseries Lemma}[section]
\newtheorem{assumption}{\rm\bfseries Assumption}[section]
\newtheorem{proposition}{\rm\bfseries Proposition}[section]

\renewcommand{\t}{^{\mbox{\tiny\sf T}}}

\begin{document}

\title{Ultrafast Synchronization via Local Observation}
\author{Hai-Tao Zhang$^{1}$}
\author{Ming-Can Fan$^{2}$}
\author{Yue Wu$^1$}
\author{Jianxi Gao$^3$}
\author{ H. Eugene Stanley$^4$}
\author{Tao Zhou$^5$}\email{zhutou@ustc.edu}
\author{Ye Yuan$^1$}\email{yye@hust.edu.cn}
\affiliation{$^1$School of Automation, Huazhong University of Science
  and Technology, Wuhan 430074, P.R. China\\
$^2$Department of Mathematics, Huizhou University, Huizhou 516007,
  Guangdong, P.R. China \\
$^3$Department of Computer Science, Rensselaer Polytechnic Institute,
  Troy, New York, 12180, USA\\
$^4$Center for Polymer Studies and Department of Physics, Boston
  University, Boston, MA 02215, USA\\
$^5$ CompleX Lab, Web Sciences Center, University of Electronic Science
  and Technology of China, Chengdu 611731, P.R. China
}

%\date{}% It is always \today, today,

\begin{abstract}

Rapid expansions of their size and frequent changes of their topology
make it difficult to observe and analyze complex networks. We explore
the properties of the Hankel matrix and propose an algorithm for
calculating the final synchronization state that uses a local
observation of a single node for a time period significantly shorter
than the synchronization process. We find that synchronization can be
achieved more quickly than the routine rhythm. This finding refines our
understanding of the abundant ultrafast synchronization phenomena
observed in nature, and it enables the efficient design of self-aligned
robots.

\end{abstract}


\maketitle

Synchronization is ubiquitous in nature \cite{Buck1988,Chen2017},
man-made systems \cite{Kapitaniak2012}, and patterns of human behavior
\cite{Neda2000}.  Understanding synchronizing processes and strategies
for enhancing and depressing synchronizability have already benefited
biological and engineering systems \cite{Strogatz2003}, including
foraging \cite{Parris1999}, predator avoidance \cite{Ioannou2012},
migration \cite{Weimerskirch2001}, collective control of unmanned air
vehicles \cite{Augugliaro2014}, and the self-organized formation of
multi-robot systems \cite{Rubenstein2014}.  Synchronization phenomena
and such closely related concepts as collective motion and consensus
have already attracted the attention of researchers in many branches of
science \cite{Arenas2008,Vicsek2012}.

Many mechanisms have been proposed to explain synchronization phenomena
\cite{Vicsek2012}. The best known is the neighborhood coordination
mechanism \cite{Vicsek1995} in which the activity of each individual is
affected by their nearest neighbors. The neighbors of an individual are
defined to be (i) those inside a ball-shaped range of a fixed radius
\cite{Vicsek1995,Couzin2002}, (ii) those directly connected in a network
\cite{Barahona2002,Nishikawa2003}, or (iii) those, limited in number,
that are closest \cite{Read2011}. A ``hierarchical leadership model''
was proposed \cite{Nagy2010} to explain the flock activity of
pigeons. Each pigeon follows its leader and is in turn followed by other
pigeons, resulting in a hierarchical leader-follower network without
directed circles.

Empirical studies have found that synchronization emerges quickly in
real-world ecological and biological systems
\cite{Buhl2006,Couzin2007,Attanasi2014}. In contrast, the
synchronization produced by the neighborhood coordination mechanism is
gradual. Although many methods have been proposed to speed up the
synchronizing process---e.g., adjusting the range of the neighborhood
\cite{Gao2011}, introducing individual adaptive speeds \cite{Zhang2009},
and optimizing the strength of interactions \cite{Gao2010}---the
neighborhood coordination mechanism cannot achieve the extremely rapid
synchronization observed in real-world systems. The hierarchical
leadership model has also not been validated in large-scale systems and
has been challenged by an in-depth analysis that finds no acyclic
structure \cite{Xu2012}. Two candidate mechanisms, information
propagation \cite{Attanasi2014,Cavagna2015} and predictive protocol
\cite{Zhang2008,Zhang2008b,Zhang2008c}, have been proposed to explain
ultrafast synchronization. The former argues that direction change
information can quickly propagate throughout the flock without
attenuation, and the latter that an individual, such as a bird or a
fish, is able to predict the near-future moving trajectories of
neighbors, and thus is more able to anticipate collective
motion. Understanding ultrafast synchronization is still an open
challenge, however, because these two proposed mechanisms need further
experimental validation. In addition, it is probable that the observed
phenomena are the result of the integrated effects of multiple
mechanisms.

We here propose an alternative mechanism for ultrafast network
synchronization. In connected networks with first-order linear dynamics
\cite{Saber2004}, we find that the record of the past states of a single
node can be used to achieve ultrafast synchronization.  Monitoring
additional nodes in the neighborhood of the initial node further
accelerates the synchronization. We demonstrate the ultrafast
synchronizing speed of this mechanism using simulations of
representative network models and of a variety of real networks.

In an $N$-node directed network, when there is an edge from node~$j$ to
node~$i$ there is an entry $a_{ij}=1$ in the adjacency matrix $\mathcal
A$. If not, then $a_{ij}=0$. The state $x_i$ of an arbitrary node~$i$
follows a discrete-time linear dynamics
\begin{equation}
x_i(t+1)=x_i(t)+\epsilon \sum_{j=1}^N  a_{ij}\left[ x_j(t)-x_i(t)
  \right],
\end{equation}
where $\epsilon$ is the sampling period, which is small enough
($\epsilon\leq 1/d_{\max}$, $d_{\max}$ being the maximal out-degree)
to guarantee convergence \cite{Saber2004}. Then the dynamics of the
entire network is
\begin{eqnarray}
x(t+1)=P x(t),
\label{olfsteps}
\end{eqnarray}
where $x=(x_1,x_2,\cdots,x_N)\t$, $P=I- \epsilon
(\mathcal{D}-\mathcal{A})$, $I$ is the unit matrix, and $\mathcal
D=\text{diag}\{{\bf 1}^T\mathcal{A}\}$ where ${\bf 1}$ is an
$N$-dimensional all-$1$ vector. The state $x(t)$ asymptotically
converges to the final value $x(\infty)=\mu x(0){\bf 1}$ if the spectral
radius of $P$ is no greater than $1$. Here $\mu$ is the left eigenvector
of $P$ corresponding to eigenvalue $1$, which also satisfies the
normalization condition $\mu {\bf 1}=1$. Specially, for undirected
networks or balanced directed networks (i.e., $\sum_j a_{ij}=\sum_j
a_{ji}$ for every node $i$), $x(\infty)=\frac{1}{N}\sum_{i=1}^N
x_i(0){\bf 1}$.

We designate node $i$ to be the node from which we gather
time-sequential information about itself and about $\ell$ neighboring
nodes $i_1,\cdots,i_\ell$ (``monitored nodes'') as
$y_i=(x_i,x_{i_1},\cdots, x_{i_{\ell}})\t$. We define an $(L+1)\times N$
matrix $C_i$ in which column $i$ in the first row and columns $i_j$ in
$(j+1)$th rows are 1 $(j=1,2,\cdots,l)$. All other elements are 0. Thus
\begin{equation}
y_i(t)=(x_i(t),x_{i_1}(t),\cdots, x_{i_{\ell}}(t))\t = C_ix(t).
\end{equation}
We designate $D_i$ to be the smallest integer that satisfies condition
$C_iq_i(P)=0$, where $q_i$ is a monic polynomial with a degree $D_i+1$,
\begin{equation}
q_i(z)= z^{D_i+1}+\sum_{j=0}^{D_i}\alpha^{(i)}_j z^j,
\end{equation}
where $\alpha^{(i)}_j$ are free parameters. Since
$y_i(t+j)=C_ix(t+j)=C_iP^jx(t)$, for any $t$ we have
\begin{equation}
\sum_{j=0}^{D_i+1} \alpha^{(i)}_{j} y_i(t+j)=
\sum_{j=0}^{D_i+1}C_iq_i(P)x(t)=0,
\end{equation}
where $\alpha^{(i)}_{D_i+1}=1$.

Focusing on the Z-transform $Y_i(z) = \mathcal{Z}(y_i(t))$, from Eq.~(5)
and the time-shift property of the Z-transform we have
\begin{equation}
Y_i(z) = \frac{\sum_{j=1}^{D_i+1} \alpha^{(i)}_j \left(\sum_{h=0}^{j-1}
  y_i(h) z^{j-h}\right)}{q_{i}(z)} \triangleq \frac{H(z)}{q_{i}(z)}.
\end{equation}
Note that according to the definition of $P$ in \eqref{olfsteps}
\cite{ye} the only unstable root of $q_i(z)$ is the one at $1$. We then
define
\begin{equation}
p_i(z) = \frac{q_i(z)}{z-1} = \sum_{j=0}^{D_i} \beta_j z^j.
\label{eq:p_polynomial}
\end{equation}
From Eq.~\eqref{eq:p_polynomial} we deduce that
{\small\begin{equation}
\sum_{j=0}^{D_i} \beta_{j} \left(y_i(t+j+1)- y_i(t+j)\right)=
\sum_{j=0}^{D_i} \beta_{j}C_i(P-I)P^jx(t)=0.
\label{eq:q_polynomial}
\end{equation}}

Using the final value theorem in \eqref{eq:p_polynomial} and some simple
algebra we find the consensus value $\phi \mathbf{1}$,
\begin{equation}
\label{eq:consensus_final}
\phi \mathbf{1}= \lim_{z \to 1} (z-1) Y_i(z) =
\frac{H(1)}{p_i(1)}=\frac{y_{D_i}^T \beta}{\mathbf{1}^T \beta},
\end{equation}
where $y_{D_i}^T = \begin{bmatrix} y_i(0) & y_i(1) & \ldots &
  y_i(D_i) \end{bmatrix}$ and $\beta_{(D_i+1) \times 1}$ is the vector
of coefficients of the polynomial $p_i(z)$ defined in
Eq.~\eqref{eq:p_polynomial}.

We denote the Hankel matrix
\begin{eqnarray*}
&\Gamma\left\{y_i(0), y_i(1), \ldots, y_i\left(\left\lceil
  \frac{k+1}{\ell+1} \right\rceil+k-1\right)\right\}\\
 &=\left[ \begin{array}{cccc} y_i(0) & y_i(1) &\ldots
&y_i(k)\\ y_i(1) & y_i(2) &\ldots & y_i(k+1) \\ \vdots
&\vdots &\ldots &\vdots \\ y_i(\lceil \frac{k+1}{\ell+1} \rceil-1 )
&y_i(\lceil \frac{k+1}{\ell+1} \rceil) &\ldots &y_i(\lceil
      \frac{k+1}{\ell+1} \rceil+k-1)\end{array} \right].
\end{eqnarray*}
Node~$i$ then stores $y_i(t)$ $(t=0, 1, \ldots)$ in memory and
recursively builds the Hankel matrix $H_{i,\ell}^k$,
\begin{widetext}
\begin{equation}
H_{i,\ell}^k=\Gamma\left\{y_i(1) - y_i(0), y_i(2)-y_i(1), \ldots,
y_i\left(\left\lceil \frac{k+1}{\ell+1} \right\rceil+k\right)
-y_i\left(\left\lceil \frac{k+1}{\ell+1}
\right\rceil+k-1\right)\right\},
\end{equation}
\end{widetext}

where $\lceil x \rceil$ is the nearest integer not less than $x$, and
$H_{i,\ell}^k$ always has more rows than columns. Node $i$ then
calculates the rank of $H_{i,\ell}^k$ and increases the dimension $k$
until $H_{i,\ell}^k$ loses column rank and stores the first defective
Hankel matrix $H_{i,\ell}^K$. Here $K$ is a good estimation of $D_i$
\cite{ye}. Node $i$ then calculates the normalized kernel
$\beta=(\beta_0,\cdots,\beta_{K-1},1)\t$ of $H_{i,\ell}^K$, i.e.,
$H_{i,\ell}^K\beta ={\bf 0}$ according to
Eq.~\eqref{eq:q_polynomial}. Once $\beta$ is obtained and combined with
the previously memorized $[y_i(0) \;\;y_i(1) \;\;\ldots \;\; y_i(K)]$,
node $i$ can use the final value theorem in
Eq.~\eqref{eq:consensus_final} and calculate the final global
synchronized value
\begin{equation}
\phi {\textbf1}=\frac{[y_i(0)  \;\;y_i(1) \;\;\ldots \;\; y_i(K)]
  \beta}{\textbf{1}\t  \beta},
\end{equation}
where $\phi=\mu x(0)$.

We calculate the global synchronized value $\phi$ using node $i$ within
$N_{i,\ell}=\lceil \frac{K+1}{\ell+1} \rceil+K$ iteration steps from the
proposed algorithm. All nodes then propel themselves toward the
calculated destination $\phi$. Given the observed node $i$ and its
$\ell$ monitored nodes, the synchronizing time of the method can thus be
quantified using $N_{i,\ell}$. To quantify the synchronization speed of
the routine process, we directly simulate the dynamics \eqref{olfsteps}
and define the minimal convergence steps $M$ as when the state
difference of all node pairs, e.g., $\sum_{i>j} |x_i-x_j|$, drops below
a small threshold $\delta$ (here we set $\delta=10^{-3}$). The smaller
the value of $M$, the more rapid the synchronization.

\begin{table}[!htb]
\caption{AMML and AMCS of 100 ER, BA and WS networks with size
  $N=100$. The initial state of each node is randomly selected in the
  range $[-2,2]$. The rewiring probability of WS networks is set as
  $p=0.9$.}
 \begin{tabular}{ p{2cm} p{1cm} p{1cm} p{1cm} p{1cm} p{1cm} p{1cm}}
% \hline
%&MML  &MML &MML &MML  &MCS \\
%&($\ell=0$)&($\ell=1$) &($\ell=2$) &($\ell=4$) &\\
%\hline
%Regular &1.6 &8.064 & 8.284   &7.436  &18\\
%%Complete &1 &1 &1 &1 &1 \\
 \hline
&AMML &AMML &AMML &AMML &AMCS \\
&($\ell=0$)&($\ell=1$) &($\ell=2$) &($\ell=4$) &\\
\hline
ER($\rho=0.1$) &52.34 &44.63 &35.80 &33.85   &281.51\\
%ER($p=0.15$) &20.90 &13.29 &11.86 &10.53   &118.06\\
ER($\rho=0.2$) &18.92 &11.14&10.21 &9.29   &74.61\\
BA($m=3$) &77.17 &67.22 &62.86  &63.92  & 507.39\\
%BA($m=4$) &55.30 &38.01  &34.98  &34.47  &320.80\\
BA($m=5$) &44.49 &26.18  &23.73  &22.79 &233.22\\
WS($z=6$) &78.22 &69.34 &62.93  &46.78  &582.31\\
%WS($m=8$) &51.88 &34.03 &28.85  &24.88   &333.10\\
WS($z=10$) &37.83 &19.86 &16.54  &13.98  &233.20\\
\hline
\end{tabular}
\label{tab: recoersfws}
 \end{table}

\begin{figure}[!htb]
\centering	
\includegraphics[width=\linewidth]{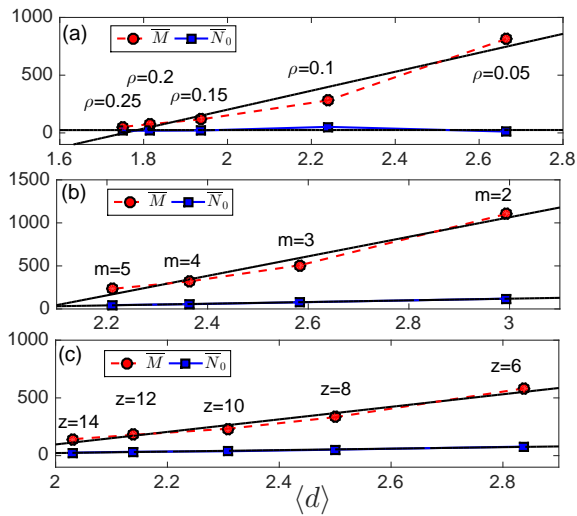}
\caption{The synchronization time, $\overline{M}$ and $\overline{N}_0$,
  versus the average distance $\langle d\rangle$ for ER (a), BA (b) and
  WS (c) networks. The black lines represent the linear fitting for data
  points by the least squares estimation. The network size is set as
  $N=100$, and the rewiring probability is set as $p=0.9$ for WS
  networks.}
\label{aplersfws}
\end{figure}

We consider three types of undirected network model, i.e., the
Erd$\ddot{o}$s-R$\acute{e}$nyi (ER) \cite{er60pub}, the
Barab\'asi-Albert (BA) \cite{baal99}, and the Watts-Strogatz (WS)
\cite{wa98}. In an ER network, node pairs are connected with a
probability $\rho$. Initially a BA network is a small clique of $m$
nodes, and at each time step a single node is added with $m$ edges
connecting to existing nodes. The probability of selecting an existing
node is proportional to its degree. Initially a WS network is a
one-dimensional lattice in which each node connects to $z$ neighbors,
and each edge then has a constant probability $p$ of being rewired. The
average degree of an BA network is approximately $2m$, and the average
degree of an WS network is $z$. We generate 100 networks of size $N=100$
for each model. In each network, each node has a single chance of being
chosen, and we then independently pick up its $\ell$ neighbors for 100
times. Both the average minimal memory length (average MML, or AMML,
$\overline N_{\ell}$) and average minimal convergence steps (average MCS
or AMCS, $\overline M$) are obtained by averaging over all independent
runs. As shown in Table~\ref{tab: recoersfws}, even when $\ell=0$ we
know only the record of the observed node and not of the neighboring
monitored nodes, and the synchronization speed of this method is much
faster than the routine process, as indicated by how much smaller the
value of $\overline{N}_0$ is than $\overline M$. In addition,
$\overline{N}_\ell$ decreases when $\ell$ increases, suggesting that the
synchronization can be further accelerated by including the monitored
nodes.

\begin{figure}[!htb]
\centering	
\includegraphics[width=.99\linewidth]{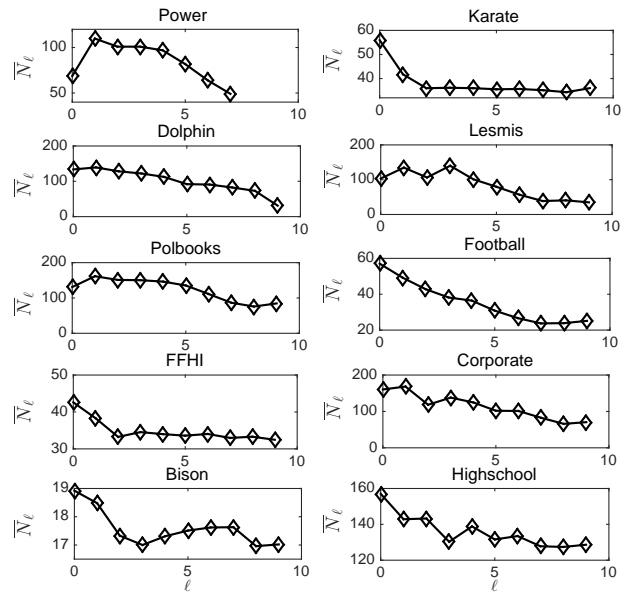}
\caption{AMML as a function of the number of monitored neighbors $l$ for
  the 10 real networks. Given $\ell$, we only select observable nodes
  with degree no less than $\ell$ to implement the simulation and then
  get average MML over these nodes. Since the maximal degree of Power is
  7, the corresponding maximal $\ell$ in the first plot is 7 as well.}
\label{minreal}
\end{figure}

The state of a node is directly affected by its neighbors, and the state
of the neighbors are in turn affected by their neighbors, i.e., the
influence spreads over edges leading to interplay among all node
pairs. According to discrete dynamics (1), the average number of steps
required for the influence from a randomly selected node to reach
another randomly selected node is equal to the average distance $\langle
d\rangle$. Thus the synchronization time is strongly dependent on the
average distance. Figure~1 shows a varying of the model parameters and
the relationship between the synchronization time and the average
distance in both our method and the routine process. The synchronization
time $\overline{M}$ required by routine method is much longer than
$\overline{N}_l$ even when $l=0$. Both relationships $\overline{M} -
\langle d\rangle$ and $\overline{N}_0 - \langle d\rangle$ approximately
fit a linear function, but the increasing rate of $\overline{M}$ is much
larger than that of $\overline{N}_0$. We thus expect that in networks
with a larger $\langle d\rangle$ the advantage enjoyed by
$\overline{N}_0$ will become even more significant.

\begin{table}[!htb]
\centering
\caption{Topological features and synchronization time for the 10 real
  networks under consideration. $N$, $E$, $\langle k \rangle$ and
  $\langle d\rangle$ represent the number of nodes, the number of edges,
  the average degree and the average distance, respectively. The former
  8 networks are undirected while the last two are directed. The average
  distance of a directed network is defined as $\langle d\rangle =
  \frac{1}{N(N-1)}\sum_{i\neq j}d(i,j)$, where $d(i,j)$ is the distance
  from node $i$ to node $j$. If the network is not strongly connected,
  $\langle d\rangle = \infty$. As clearly observed from this table,
  $\overline{N}_0$ is much smaller than $\overline{M}$, indicating the
  remarkable advantage of the present method.}
\begin{tabular}{ p{1.6cm} p{0.9cm} p{0.9cm}  p{1.2cm} p{1.2cm} p{1.2cm} p{1.2cm}}
 \hline
 &\rm{\textbf{$N$}} & \rm{\textbf{$E$}} &\rm{\textbf{$\langle k
     \rangle$}}  &\textbf{$\langle d\rangle$}
 &\rm{\textbf{$\overline{N}_0$}} &\rm{\textbf{$M$}} \\
\hline
Karate &34 &78 &  4.588   &2.408 &55.93  &501 \\
Power &57 &78  & 2.737   &4.954 &214.04 &4247\\
Dolphin & 62 & 159 &  5.129    &3.357 &133.61 &1595 \\
Lesmis &77 & 254 & 6.597   &2.641 &103.33 &2328\\
Polbooks &105 &441 &8.4  &3.079  &130.94  &2285\\
Football & 115 & 613 & 10.661   &2.508 &57.04 &527\\
FFHI & 180 & 2239 &24.667   &  2.148  &49.68 &621\\
Corporate &197 &801 &8.132   &2.106  &160.36  &3712 \\
Bison &26 &314 &12.0769  &1.571  &18.92  &161\\
Highschool &70 &366 &5.229 &$\infty$ &157 &768\\
\hline
\end{tabular}
\label{tab: realnet}
 \end{table}

\begin{figure}[htb]
\centering	
\includegraphics[width=8.5cm]{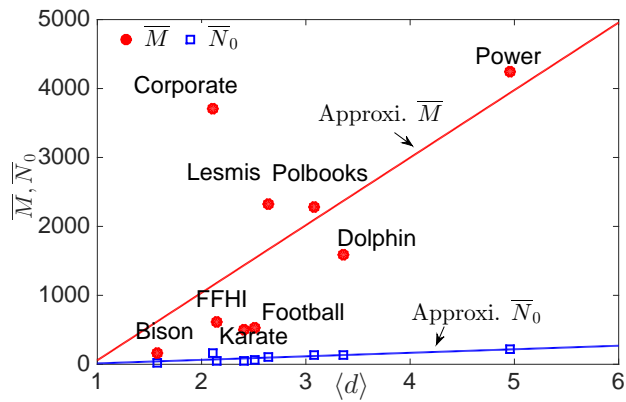}
\caption{The synchronization time, $\overline{M}$ and $\overline{N}_0$,
  versus the average distance $\langle d\rangle$ for the 9 real networks
  except Highschool whose average distance is infinite. The red circles
  and blue squares denote the synchronization times by the routine
  procedure and the present method, respectively. The red and blue lines
  represent the linear fittings for data points by the least squares
  estimation.}
\label{aplma}
\end{figure}

We next examine ten real-world networks:

\begin{itemize}

\item[{(i)}] ``Karate'': A karate network that comprises the friendships
  among 34 members of a karate club at a US university \cite{zww77jar}.

\item[{(ii)}] ``Power'': Data on a US electric power network from the
  early 1960s, which we acquired from a 57-bus case download
  from~\cite{cr93url}.

\item[{(iii)}] ``Dolphin'': An undirected social network of frequent
  contacts among 62 bottlenose dolphins in a community at Doubtful
  Sound, New Zealand \cite{ld03bes}.

\item[{(iv)}] ``Lesmis'': The network of fictional characters in Victor
  Hugo's novel {\it Les Miserables}, where an edge denotes the
  co-appearance of the two corresponding characters \cite{kde93}.

\item[{(v)}] ``Polbooks'': The network of books about recent US politics
  in Amazon.com, where the edges between books represent the frequent
  co-purchasing of books by the same buyer \cite{kva08}.

\item[{(vi)}] ``Football'': The network of American faootball games
  between Division IA colleges during the regular season in Fall 2000,
  where each node represents a team and two teams are connected if they
  have regular seasonal games \cite{gm02pnas}.

\item[{(vii)}] ``FFHI'': The face-to-face human interaction network
  where each node denotes an individual in a school
  \cite{Starnini2013}.

\item[{(viii)}] ``Corporate'': A European corporate community network in
  which nodes represent firms and two firms are connected if they share
  at least one manager or director \cite{kbm12}.

\end{itemize}

\noindent
These eight networks are undirected. To verify the generality of the
results, we also examine two directed networks.

\begin{itemize}

\item[{(ix)}] ``Bison'': Is the dominance relationships among American
  bison in 1972 on the National Bison Range in Moiese, Montana, where
  each node denotes a bison and a directed edge from node $i$ to node
  $j$ represents the dominance relationship: $i$ over $j$.

\item[{(x)}] ``Highschool'': The network of friendships among boys in a
  small high school in Illinois \cite{coleman64}. Highschool is directed
  because student $i$ can identify $j$ as a friend even when $j$ does
  not identify $i$ as a friend.

\end{itemize}

\noindent
Table~\ref{tab: realnet} provides the structural statistics and
synchronization time of the ten real-world networks. When we compare the
last two columns we see that our method produces much faster
synchronization. In most cases our method is ten times faster than the
routine procedure even without using monitored nodes.
Figure~\ref{minreal} shows that despite some fluctuations the
synchronization time $\overline{N}_\ell$ further decreases as $\ell$
increases, indicating additional benefit when monitored nodes are
introduced. Similar to that shown in Fig.~\ref{aplersfws}, both
$\overline{N}_0$ and $\overline{M}$ increase with $\langle
d\rangle$. Figure~\ref{aplma} shows linear fittings for visual guidance,
where it is clear that the increasing rate of $\overline{M}$ is much
larger than $\overline{N}_0$. Thus experimental analyses of disparate
real-world networks once again demonstrate the results obtained from
network models, i.e., (i) our method speeds up synchronization, (ii)
monitoring more neighbors further accelerates synchronization, and (iii)
the synchronization time is positively correlated with $\langle
d\rangle$, while the present method grows more slowly.

We have found a mechanism that leads to the ultrafast synchronization of
general networked discrete-time linear dynamics and that requires only
the historical dynamical trajectory of one observable node. In a
networked dynamical system, the state of a node is directly affected by
its neighbors, who are directed affected by their neighbors, and so
on. Thus the state of a node will affect and be affected by all other
nodes after a sufficiently long period of time. Our major contribution
here is successfully realizing this theoretical possibility by applying
Hankel matrix analysis.

Compared to the information propagation \cite{Attanasi2014,Cavagna2015}
and predictive protocol \cite{Zhang2008,Zhang2008b,Zhang2008c}, the our
proposed mechanism requires little intelligence from most individuals
but a higher level of intelligence from the observable node. This
includes both the memory to store the historical dynamical trajectory
and an ability to analyze this trajectory. In a biological system, it is
unlikely that a leader would use a Hankel matrix-like process to
determine in advance a travel direction in which to lead the flock. It
is more likely individuals with a short memory would not use the
elaborately designed Hankel matrix but instead use recent dynamical
records to acquire approximate synchronization. Thus our proposed
mechanism is also a candidate for achieving ultrafast synchronization.

We believe that this mechanism will have significant applications in
engineering systems. A group with one super leader is unlikely in the
biological world but easy to design and implement in the
humanly-constructed world. A distributed sensor network in which each
sensor communicates and interacts with its neighbors must be able to
align and move together in such scenarios as field investigation or
battleground detection. Our proposed mechanism does not require a large
number of low-intellegence sensors but only one sensor with high
intelligence.  Modern information technology (in particular, the rapid
development of intelligent hardware) allows us to produce a smart sensor
with a sufficiently long memory and the ability to analyze a Hankel
matrix. Thus this smart sensor could predict the future global state of
any kind of networked linear dynamics and shape the consensus of the
entire sensor group.

This work was partially supported by the National Natural Science
Foundation of China under Grant Nos. 61673189, 61433014, 61703175 and
51535004.

\end{document}